# Long-Term Latency Measurement of Deployed Fiber

**Florian Azendorf [1], Annika Dochhan[1], Ralf-Peter Braun[2], Michael Eiselt[1]**

[1] *ADVA Optical Networking SE, Maerzenquelle 1-3, 98617 Meiningen, Germany*

[2] *Deutsche Telekom AG, Winterfeldtstraße 21, 10781 Berlin, Germany*

fazendorf@advaoptical.com

**Abstract:** Using a Correlation-OTDR we measured the latency of fibers in a deployed cable and calculated the time coefficient of the fiber temperature changes. Annual temperature variations of 25K were estimated for the deployed fiber.

**OCIS codes:** (060.2270) Fiber Characterization; (120.4825) Optical Time Domain Reflectometry.

## 1. Introduction

Latency is becoming a critical parameter in future 5G networks. Synchronization protocols require a stable and symmetric latency between master and slave clocks. Other applications, like the transmission of radio phase array signals, as investigated in the European BlueSpace project [1], require a very low differential latency between different optical paths. One of the factors strongly impacting latency is temperature, which mainly impacts the refractive fiber index, resulting in a latency change of approximately 6 ppm/K [2, 3]. In addition, strain effects in a jumper cable with a tight buffer can lead to a temperature delay coefficient (TDC) of 17 ppm/K, as measured in a laboratory environment [4]. So far, to our knowledge, the latency changes of a fiber deployed in the ground has not been measured with a high accuracy. In this paper, we report on those measurements, using a Correlation-OTDR (C-OTDR), yielding a high accuracy of the absolute fiber latency on the order of a few picoseconds [5].

## 2. Experimental setup

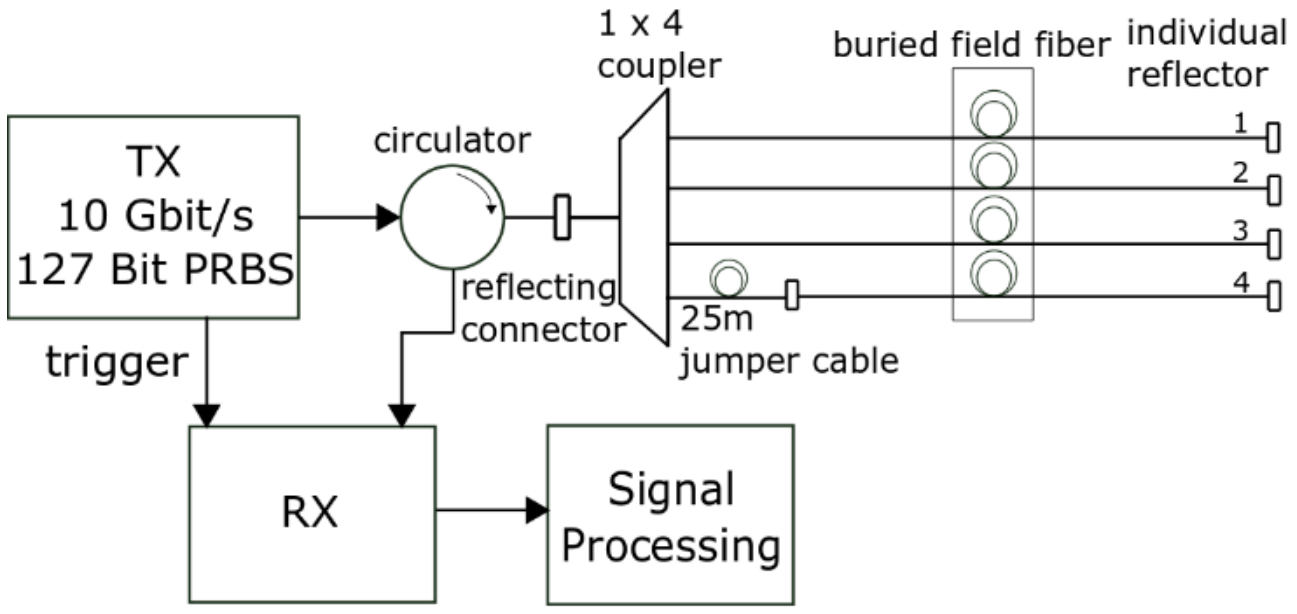

**Fig. 1:** Measurement setup for simultaneous latency characterization of four deployed fibers.

We characterized four fibers in a deployed 8.5-km underground cable between the ADVA office and a Deutsche Telekom central office in Meiningen, in Central Germany. The setup is shown in Fig. 1. A laser at a wavelength of 1550 nm was modulated using a Mach-Zehnder modulator with a 10-Gbit/s 127-bit PRBS burst followed by zeros to fill a 100-µs packet, sufficient to measure a 10-km fiber. The signal was sent via a circulator and passed a connector with an air gap, providing a reference reflection from the near end of the fiber. To measure all fibers simultaneously, the signal was split using a 1x4-coupler. Fiber loops using 1x2-couplers provided a high reflection from the far end of each fiber (individual reflectors). The reflected light was received after the circulator and, after a PIN/TIA receiver, observed on a 50-GS/s real time oscilloscope. The real time oscilloscope was synchronized with the burst trigger from the PRBS source, and 2000 traces were recorded and averaged. The averaged signal is shown in Fig. 2a, showing the PRBS sequences reflected from the four fiber ends. Note that the first two reflected packets are overlapping. The averaged signal was correlated with the transmitted PRBS and filtered to eliminate pre- and post-cursors in the correlation function due to the isolated PRBS. The correlated signal is shown in Fig. 2b. As the lengths of two fibers matched within less than 100 ps such that their end reflections were indistinguishable, we added a 25-m jumper cable to the 4$^{th}$ fiber. The resulting higher round-trip latency of 250 ns can be seen in Fig. 2b. As the 50 GS/s oscilloscope yielded only a sample resolution of 20 ps, a raised cosine function was fitted to the highest correlation peaks to improve the position accuracy to a few picoseconds [5]. The cosine fit is show in Fig. 2c.

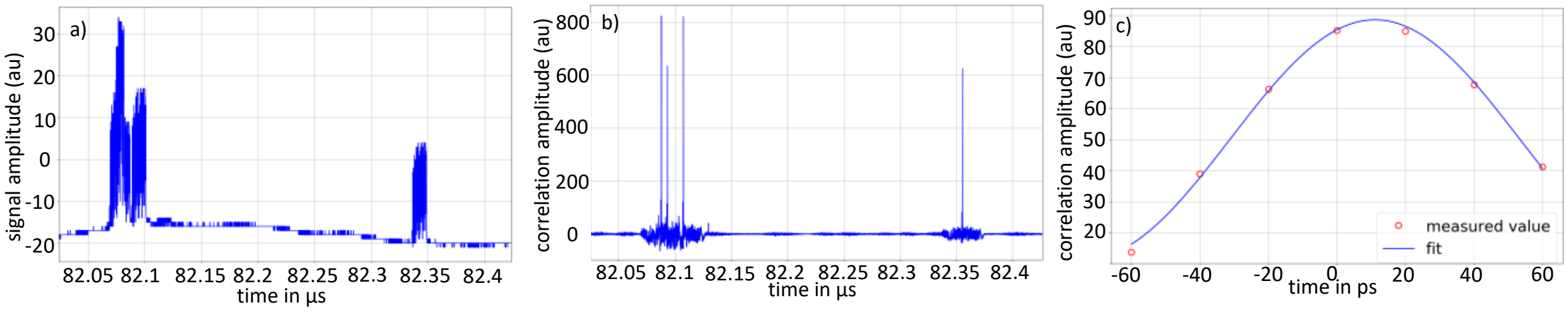

**Fig. 2:** a) Averaged signal showing reflected PRBS sequences; b) averaged signal correlated with transmitted PRBS sequence showing reflection peaks from four fiber ends; c) fit of one correlation peak to a raised cosine function.

## 3. Results

Using the C-OTDR, the round-trip latency of the four fibers was monitored over approximately two weeks with a half day interrupted around the 8th day. As shown in the blue curve in Fig. 3a, the maximum latency increase over the time, referenced to the latency at the start of the measurement, was 800 ps. Fig. 3a also contains the outside air temperature, as obtained from a data base maintained by the German Meteorological Service [7]. It can be seen that the latency does not immediately follow the temperature but is dampened and has some time delay. Similar behavior was reported in [6], where the temperature in soil in different depths was measured. We modelled the temperature variation of the fiber as a low-pass filtered response to the outside temperature, as reported in Section 4 of this paper. This delayed response of the fiber temperature led to daily latency variations of about 200 ps. All fibers exhibit similar behavior, as shown in detail in Fig. 3b for a 12-hour period. However, fiber 4 shows a larger latency variation, which we attribute to the additional 25 m of jumper cable, which was subjected to the laboratory environment, resulting in a round-trip latency variation of approximately 10 ps/K [2]. The latency difference of 20 ps corresponds to a temperature increase of 2 K, which is a typical value in our laboratory over the day.

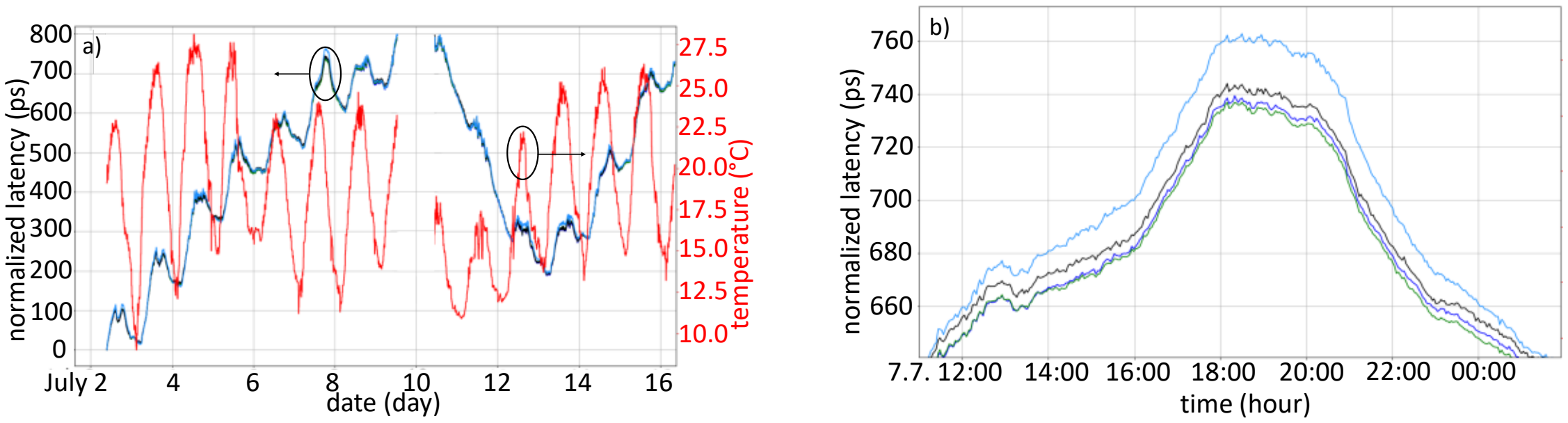

**Fig. 3**: Results of the long-term measurements; a.) relative round-trip latency of four deployed fibers and outside temperature over two weeks in July 2018 with a half day interruption from 9th to 10th; b.) latency evolution of four fibers on July 7th .

Another important parameter for synchronization applications is the latency difference (or skew) between the fibers. Fig. 4 shows the skew variations between fibers 1, 2, and 3 over the measurement period, offset by the minimum skew. The maximum variation was approximately 12 ps, seen between fibers 1 and 3. It can also be seen that the skew variation approximately tracks the absolute latency changes (blue curve in Fig. 3a) and might be attributed to TDC differences between the fibers on the order of 1%.

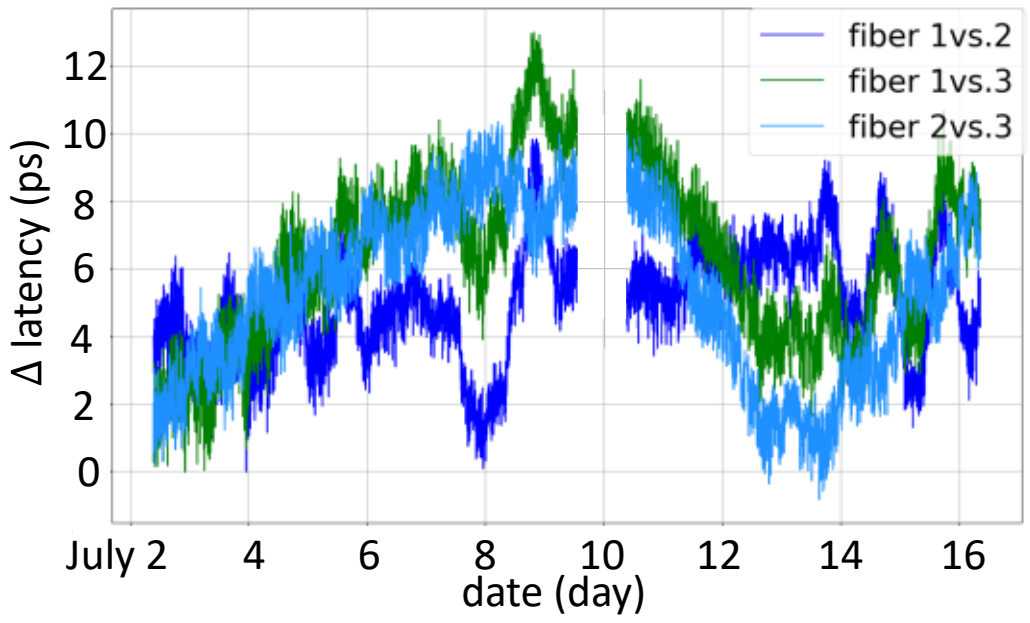

**Fig. 4:** Variation over time of differential latency (skew) between fibers 1, 2, and 3.

## 4. Modelling of the fiber temperature

We modelled the evolution of the temperature of the fiber deployed in the ground as a 1st order low-pass filtered function of the outside air temperature. The low-pass filter parameters were derived by a fit of the filtered temperature to the measured latency variations of the fiber. This assumes that the latency variations are linear with the fiber temperature, which was established e.g. in [4]. Fig. 5a shows the measured latency variation as well as the filtered temperature changes with fitting parameters of 7.5 ppm/K for the TDC and 12.7 days for the filter time constant. It can be seen that the modelling replicates the slower trends of latency changes over multiple days as well as some of the smaller day-night variations. The peak-to-peak temperature variation of the cabled fiber was only 1.5 K during the two weeks of measurement. Using this model for the annual variations of the fiber temperature based on data of the air temperature obtained from the German Meteorological Service [7], we estimated the variation of the fiber temperature over the year. Fig. 5b shows the air temperature variation with a resolution of 10 minutes and the resulting filtered fiber temperature with a peak-to-peak variation of approximately 28 K. This temperature variation would lead to a round-trip latency variation of the 8.5-km fiber link by approximately 17 ns over the year. A skew variation of approximately 200 ps can be estimated based on the results in Figure 4.

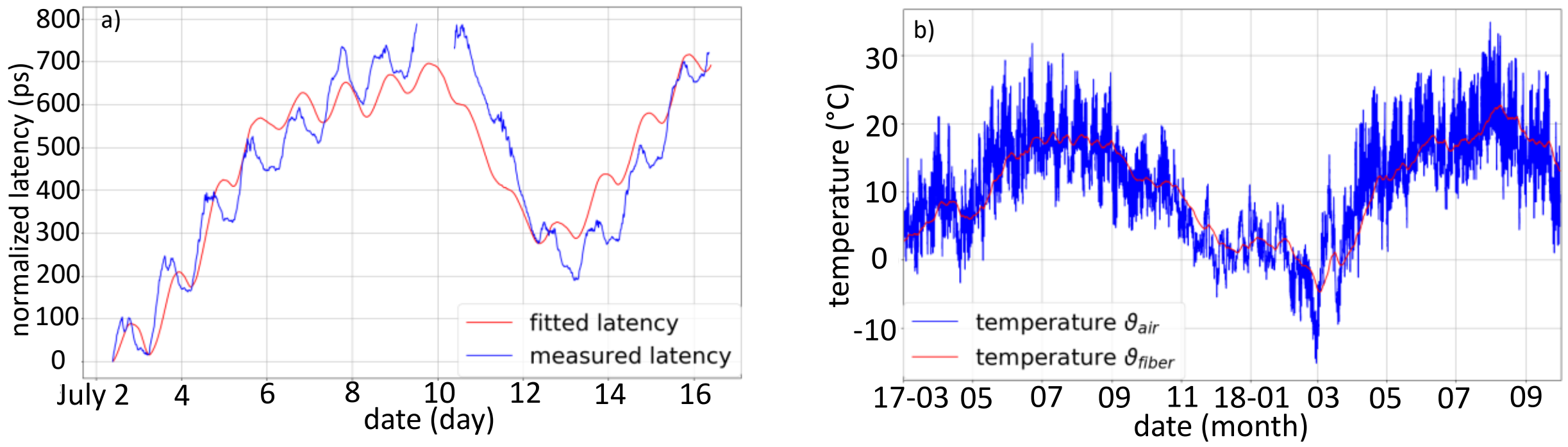

**Fig. 5:** Modelling of cabled fiber temperature a) Measured fiber latency(blue) and latency based on filtered cable temperature (red); b) Temperature evolution over 18 months: 10-min air temperature data(blue) and slower changing fiber temperature (red).

## 5. Summary

Using a correlation OTDR, with an accuracy of approximately 2 ps, we measured the latency variations of four fibers in a deployed 8.5-km cable. Over a period of 14 days in the summer, a maximum round-trip latency variation of 800 ps was measured. The measured skew variations between the fibers were up to 12 ps. Based on these measurements, a time constant of 12.7 days for the temperature change of the fibers in the cables was extracted, which would lead to a temperature change of 28 K over the year, resulting in variations of latency and skew of 17 ns and 200 ps, respectively.

## 6. Acknowledgment

This project has received funding from the European Union´s Horizon 2020 research and innovation programme under grant agreement No 762055 (BlueSpace Project).